\documentclass[final]{svjour2}
\usepackage{graphicx}
\usepackage{rotating}
\usepackage{amssymb}
\usepackage{mathptmx}
\usepackage[numbers]{natbib}
\makeatletter
\journalname{Journal of Low Temperature Physics}

\bibpunct{}{}{,}{s}{}{,}

\begin{document}

\newcommand{\hdblarrow}{H\makebox[0.9ex][l]{$\downdownarrows$}-}
\title{ Investigating metastable hcp solid helium below its melting pressure}

\author{F.~Souris \and J.~Grucker \and J.~Dupont-Roc \and Ph.~Jacquier}

\institute{Laboratoire Kastler Brossel, ENS/CNRS/Universit\'e Paris 6,\\ 24 rue Lhomond \\ Paris,
F75005, France\\ Tel.:331 44323421\\ Fax:331 44323434\\ \email{fabien.souris@lkb.ens.fr}}

\date{12.06.2010}
\maketitle 

\begin{abstract}
We report a first attempt to produce metastable hcp solid helium below its melting pressure. A focused sound
pulse is emitted along the $c$-axis of a mono-domain hcp helium-4 crystal starting from a static
pressure just above the melting pressure. The sound pulse is made as simple as possible with one negative and
one positive swing only. Density at focus is monitored by an optical interferometric method. Performed numerical simulations
show that the crystal anisotropy splits the focused wave into two separate pulses, corresponding to a longitudinal wave along the $c$-axis and a radial one perpendicular to it. The amplification factor due to
focusing remains nevertheless important. Negative pressure swings up to 0.9 bar have
been produced, crossing the static melting pressure limit. Improvements in the detection method and in the
focusing amplification are proposed.

\end{abstract}

\section{Introduction}
The hcp solid phase for helium-4 is stable only  above a minimum pressure $P_f(T)\simeq 25$~bar at low temperature. Recently it has been suggested \cite{maris2009} that  this solid phase remains metastable at much lower pressures. Based on an  extrapolation of the equation of state, the spinodal line is predicted to be at negative pressure, about -10~bar.    Beside testing this prediction, there is another interest to investigate solid helium at pressures below the melting pressure. Vacancies in this quantum solid have been studied for a long time. In particular their energy $E_v$ has been measured at different pressures\cite{simmons1989}, and also computed by Quantum Monte Carlo simulations\cite{boninsegni2006}. Although there is a significant scatter on the measured values of $E_v$, it is striking that it is a strongly decreasing function of the molar volume $V_m$. A simple linear extrapolation on the data gathered in reference \cite{simmons1989} suggests that an  increase of $V_m$ by only 5\% beyond the melting volume could bring $E_v$ near zero. Although this is a large increase for a condensed phase, this is not so for solid helium which is a very compressible material. More precisely, the maximum molar volume for stable hcp helium-4\cite{greywall1971} is $V_f = 21$~cm$^3$, the aimed value is   $V_a =22$~cm$^3$, while   the predicted value of $V_m$ at the spinodal  line is $V_s=34.7$~cm$^3$. Hence $V_a$ is within a possibly accessible range.    If so, this would bring interesting new physics for solid helium. With $E_v$ near zero, vacancies could proliferate, reach a sizable population, possibly undergo a Bose-Einstein condensation, thus realizing the Andreev scenario for supersolidity\cite{andreev1969}. Or if the vacancy-vacancy interaction is strong enough\cite{boninsegni2006,mahan2006}, it could provide another mechanism for destabilizing the solid phase before reaching $V_s$. 

Many different methods have been used to study metastable condensed phases, mostly liquid, under depression\cite{balibar2003}. Metastability cannot be  easily obtained for a solid  because the interfaces between a solid and  the container walls are disordered,  and  generally nucleates the liquid phase.   Focused   sound waves provide a way to circumvent this difficulty. Starting from a pressure $P_0$ above $P_f$, a sound pulse with an amplitude $\delta P_i$ smaller than $P_0 - P_f$ at the surface of the transducer will not bring the solid below the melting pressure at  the transducer surface. On its propagation to the focus the sound pressure can be amplified by a large amount, namely $\Omega R/\lambda_s$ ($\Omega$, solid angle of the sound beam, $R$ transducer radius,  $ \lambda_s$ sound wavelength). Hence the peak pressure $P_0 + \delta P_i\Omega R/\lambda_s$ can explore the metastable domain below $P_f$ with negative $\delta P_i$.

This article reports a first attempt to  produce metastable hcp solid helium below its melting pressure. A hemispherical transducer was used to produce a converging sound wave. The density variations  at focus are monitored by an optical interferometric method. The experimental arrangement is described in section 2.  Because the sound velocity is anisotropic, the focusing is not expected to be perfect in hcp helium. A numerical simulation was made to model the remaining amplification in this case  and to determine  the relation between the optical signal and the density at focus. Numerical predictions are discussed in section 3. Experimental results  are reported and discussed in section 4.  Possible improvements and conclusions are given in section 5.

\section{Experimental arrangement and procedure}
The experimental cell is a 4~cm stainless steel cube, with five silica windows (diam. 2.5~cm), cooled from the top by a pumped helium-4 fridge in the 1.0-1.4~K temperature range. To achieve single crystal growth, the nucleation and growth is made at constant temperature and pressure
(1.2~K and 25.5~bar). An electro-crystallization device\cite{keshishev1979} provides a unique seed which falls on the bottom of the cell. The crystal is subsequently grown from this seed. The orientation of the {\it c-}facet (most often horizontal) is easily monitored visually during the
growth process which takes place below the corresponding roughening transition 1.3~K \cite{balibar2005}. 

A hemispherical piezoelectric transducer of  6~mm  internal radius (from Channel Industries) is suspended above the middle of the cell with its axis vertical, along the crystal {\it c}-axis. Such a transducer has two main resonant modes: the thickness mode frequency is $\nu_{hf}\simeq 1$~MHz, the inner and outer surfaces vibrating in opposite directions. The breathing mode corresponds to an oscillation of the transducer radius  and has a frequency $\nu_{lf}\simeq 180$kHz. In order to simplify as much as possible the pressure wave, a single oscillation of the thickness mode was used. The driving voltage is produce by an arbitrary function generator and amplified by a RF amplifier up to a voltage ranging from 100~V to 800~V on 50~$\Omega$. The  pulse shape is designed to leave the transducer at rest at the end of the pulse.  Beside the one-cycle oscillation at 1~MHz, the exciting pulse also produces a parasitic oscillation of the low frequency breathing mode which appears after the end of the 1~MHz pulse. The sound velocity in hcp solid helium is anisotropic. For longitudinal waves, the velocity is about 540~m/s along the {\it c}-axis and 460~m/s when the wave vector is in the basal plane\cite{crepeau1972}. Hence the initially spherical wave does not remain so and breaks up into two parts: one along the {\it c}-axis, named as \lq $z$-pulse\rq~hereafter, and another radial one, named \lq $r$-pulse\rq. Both show a maximum amplitude near the center of the transducer. Their relative intensity and shape will be discussed in the next section. In order to have an optical access to the transducer center two small notches (1~mm wide, 1.5~mm high) were made on the transducer rim, along a diameter.
   
The phenomena at the focus are monitored optically. A CW-laser beam ($\lambda_o = 532$~nm) propagating along the $y$-axis is focused at the cell center with a 30~$\mu$m waist.  The laser polarization is along the vertical $z$-axis. The axis origin is taken at the transducer center. Density modulations along the beam path produced by the sound wave result in an optical phase change
\begin{equation}\label{dphi}
\delta\phi(x,z) = {2 \pi\over \lambda_o}\int_{-l/2}^{l/2} {\rm d}y \ \delta n(\sqrt{x^2 + y^2},z)
\end{equation}
where $l$ is the cell length.  The change in refractive index $\delta n$  is related to that of density $\delta\rho$ through $\delta n /(n-1) = \delta\rho /\rho$. This linear approximation of the Clausius-Mossotti relation holds for helium with a relative error less than $10^{-4}$ for $\delta\rho /\rho$ up to 0.1, well beyond the strains discussed here (less than $10^{-2}$). Because of the system cylindrical symmetry around the  $z$-axis, $ \delta n$ and  $\delta\rho$ only depend on $r=\sqrt{x^2 + y^2}$ and $z$. To measure $\delta\phi(t)$, the laser beam is split before the cell to produce a reference beam which crosses the cell in an unperturbed region and is recombined with the monitoring beam on a photo-detector, making a Jamin interferometer.  $\delta\phi(t)$ contains a low frequency part $\delta\phi_{lf}(t)$ due to the transducer breathing mode (at $\nu_{lf}$), and the interesting signal  $\delta\phi_{hf}(t)$ due to the thickness oscillation (at $\nu_{hf}$). To extract the latter from  $\delta\phi(t)$, a model of the low frequency signal $A s_{lf}(t-t_0)$ has been constructed with two adjustable parameters: its amplitude $A$ and its starting time $t_0$. These parameters are determined from a fit of  $A s_{lf}(t-t_0)$ to  $\delta\phi(t)$ filtered with a low-pass filter. Then  $\delta\phi_{hf}(t)$ is found as  $\delta\phi_{hf}(t) = \delta\phi(t) - A s_{lf}(t-t_0)$.

Extracting $\delta n(t)$ from $\delta\phi_{hf}(t)$ is more involved. This amounts to determine a characteristic length $a$ of the focused sound wave such that
\begin{equation}\label{hf}  
\delta\phi_{hf}(t,0,0) = 2\pi \delta n(t,0,0) a/\lambda_o.
\end{equation}  
For a continuous sinusoidal wave, $a=\lambda_s/2$ \cite{debye1909}. There are two ways to get $a$. If one can measure $\delta\phi_{hf}(t,x,0)$ for the whole range of $x$ where it is non-zero, one can invert formula \ref{dphi} and get $\delta n(t,x,0)$ from $\delta\phi_{hf}(t,x,0)$ by an inverse Abel transform. This is not always possible however due to the small width of the  notch. A simpler method relies on numerical simulation to get the ratio $\delta\phi_{hf}(t,0,0)/ n(t,0,0) $ and thus  $a$. In the following, results are discussed using the optical length variations $\delta L_{\rm opt} =\lambda_o \delta\phi_{hf}/(2\pi )$ as the signal instead of $\delta\phi_{hf}$.

\section{Numerical simulation of a focused spherical sound wave}
Since solid helium has a close-packed hexagonal structure (hcp), its linear elastic properties are described by five elastic constants, $c_{11}$, $c_{12}$, $c_{13}$, $c_{33}$, $c_{44}$ as defined by Musgrave\cite{musgrave1969}.
  The ratio of shear stress to shear strain in the meridian plane, called $c_{55}$ is not an independent constant in hcp crystals \cite{landau7} and is equal to $(c_{11}-c_{12})/2$. Numerical  values of the $c_{ij}$ have been taken from \cite{crepeau1971}: $c_{11}/\rho=21.2$, $c_{12}/\rho= 11.1$, $c_{13}/\rho=5.49$, $c_{33}/\rho=29.0$, and
$c_{44}/\rho= 6.52$, all in units of $10^4$~(m/s)$^2$.

The axis of the transducer being parallel to the $c$-axis of the crystal, deformations are of cylindrical symmetry. Elastic waves are thus governed by two coupled second order differential equations for the only two possible components of the displacement vector, $u_r$ and $u_z$:
\begin{eqnarray}
        \rho\ddot{u}_r &=& c_{11}\,(\frac{\partial^2 u_r}{\partial r^2} + \frac{1}{r}\frac{\partial u_r}{\partial r} - \frac{u_r}{r^2}) +
        (c_{13}+c_{55})\frac{\partial^2u_z}{\partial r\partial z} + c_{55}\frac{\partial^2u_r}{\partial z^2}\\
        \rho\ddot{u}_z &=& (c_{13}+c_{55})(\frac{\partial^2 u_r}{\partial r\partial z} + \frac{1}{r}\frac{\partial u_r}{\partial z}) +
        c_{33}\frac{\partial^2 u_z}{\partial z^2} + c_{55}\,(\frac{\partial^2 u_z}{\partial r^2} + \frac{1}{r}\frac{\partial u_z}{\partial r}) \label{uzpp}
\end{eqnarray}
($c_{44}$ plays no role in this geometry)

Special care must be taken in $r=0$: for symmetry reasons, $u_r$, $\dot u_r$ and $\ddot u_r$ must be $0$ in $r=0$. Using the l'Hospital's rule, equation \ref{uzpp} can be rewritten:
\begin{equation}
        \rho\ddot u_z = 2(c_{13}+c_{55})\frac{\partial^2 u_r}{\partial r \partial z} +  c_{33}\frac{\partial^2 u_z}{\partial z^2} + 2c_{55}\frac{\partial^2 u_z}{\partial r^2}  \quad\quad(r=0)
\end{equation}

These equations are integrated  using a finite difference method and a staggered leap frog scheme with initial conditions corresponding to an undeformed  crystal at rest. The model volume is $0 \leq r \leq 7.5$ and $-7.5
\leq z \leq 7.5$ (in mm) and is discretized on a $1024\times2048$ grid. The spatial step $\delta l$ is then small enough compared to the wavelength ($\lambda_s \simeq 0.5$~mm) that the dispersion introduced by the discretization is negligible.  The time step $\delta t$ is taken to be sufficiently small to fulfill the Courant criterion: essentially one must have $\delta l/\delta t < \lambda_s\nu_{hf}$. At each time step $t$ one then calculates a new field of displacement vectors with the help of the above mentioned differential equations as a function of  the fields at time $t-\delta t$ and $t-2\delta t$. The field is then adjusted so as to fulfill the boundary conditions: $u_r$ and $u_z$ are forced to be that of the transducer on the points  belonging to it  and  $u_z(r=0,z)$ is adjusted to the value $u_z(0,z) = (4 u_z(\delta l,z) - u_z(2\delta l,z))/3$ to ensure cancelation of its derivative $\frac{\partial u_z}{\partial r}(r=0,z)$.
The other boundaries are left free, giving rise to reflected waves, which arrive at the focus with a long enough delay to be of no concern. From the displacement vector field, the change in molar volume can be computed ($\delta V_{m}/V_{m} = \frac{\partial u_r}{\partial r} +
\frac{u_r}{r} + \frac{\partial u_z}{\partial z}$) and thus the change in refractive index  $\delta n$.
\begin{figure}[th]
\begin{center}
\includegraphics[width=0.8\textwidth]{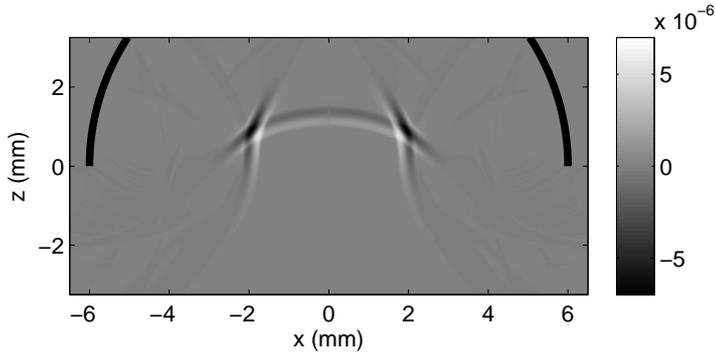}
\end{center}
\caption{Map in the $zx$-plane of the refractive index variations while the mono-oscillation sound wave is propagating. The thick black line represents the section of the  transducer surface. The initially hemispherical wave splits into an axial wave and a radial one. }\label{fig1} 
\end{figure}
An example of the wave resulting of a single oscillation of the transducer is shown on Figure \ref{fig1} at time $t =  9$~$\mu$s. As expected, the wave has split into two parts, one along the $z$-axis and a slower one which becomes almost cylindrical, traveling roughly in the $r$
direction. At the center one thus expects two pulses, as confirmed by Figure \ref{fig2}-a. When passing through the center, the $r$-wave undergoes a  Gouy phase reversal. As a result the $r$-pulse is tripolar instead of being bipolar like the $z$-pulse. Note that this is true only at the center, because only at the center does the incoming and the outgoing waves interfere.  An important result is the amplification factor.
\begin{figure}[bh]
\begin{center}
\includegraphics[width=\textwidth]{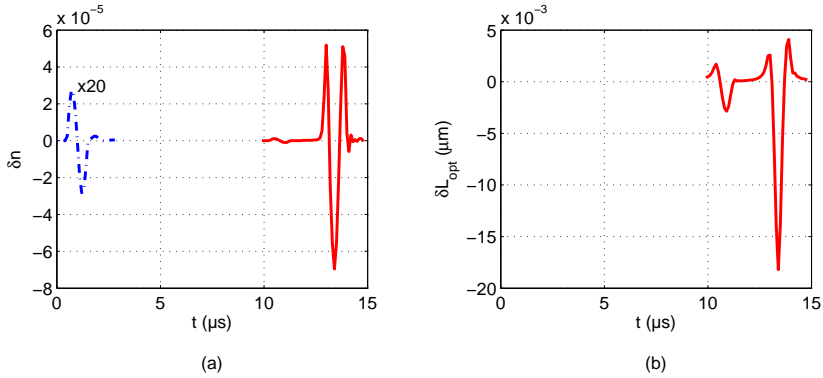}
\end{center}
\caption{(a) Refractive index modulation near the transducer surface (dashed line) and at the center (solid line). The small modulation around 11~$\mu$s is the part of the sound wave coming along the $z$-axis ($z$-pulse). The large one around 13.5~$\mu$s is the radial one ($r$-pulse). (b) Modulation of the optical length through the center of the transducer for the same pulse.}\label{fig2} 
\end{figure}
 The $z$-pulse is almost not amplified, while the $r$-pulse is amplified by a factor 50.  This is not much less than the factor 75  expected for an isotropic medium with the same $\lambda_s$.
Figure \ref{fig2} also shows the behavior of the optical length through the center of the transducer. A first striking result is that the optical length is vanishingly small for almost all times, although the integration path crosses the wave at any time. This is because the positive and the negative parts of the wave cancel almost exactly. Only when the wave reaches the center is this no more true.
The $z$-pulse seems to be magnified, but this is only a geometrical effect due to the fact
that  this part of the wave is almost plane and perpendicular to the $z$-axis. The ratio between the maximum amplitude of the optical length and the maximum amplitude of the index modulation is a characteristic length $a$ ($a \simeq 0.28$~mm).  It can be used to deduce the refractive index modulation,  which is  the quantity of interest, from the optical length, which is the measured quantity.

\section{Presentation and discussion of some  results}
\begin{figure}[thb]
\begin{center}
\mbox{
\begin{tabular}{c c}
\includegraphics[scale=0.38]{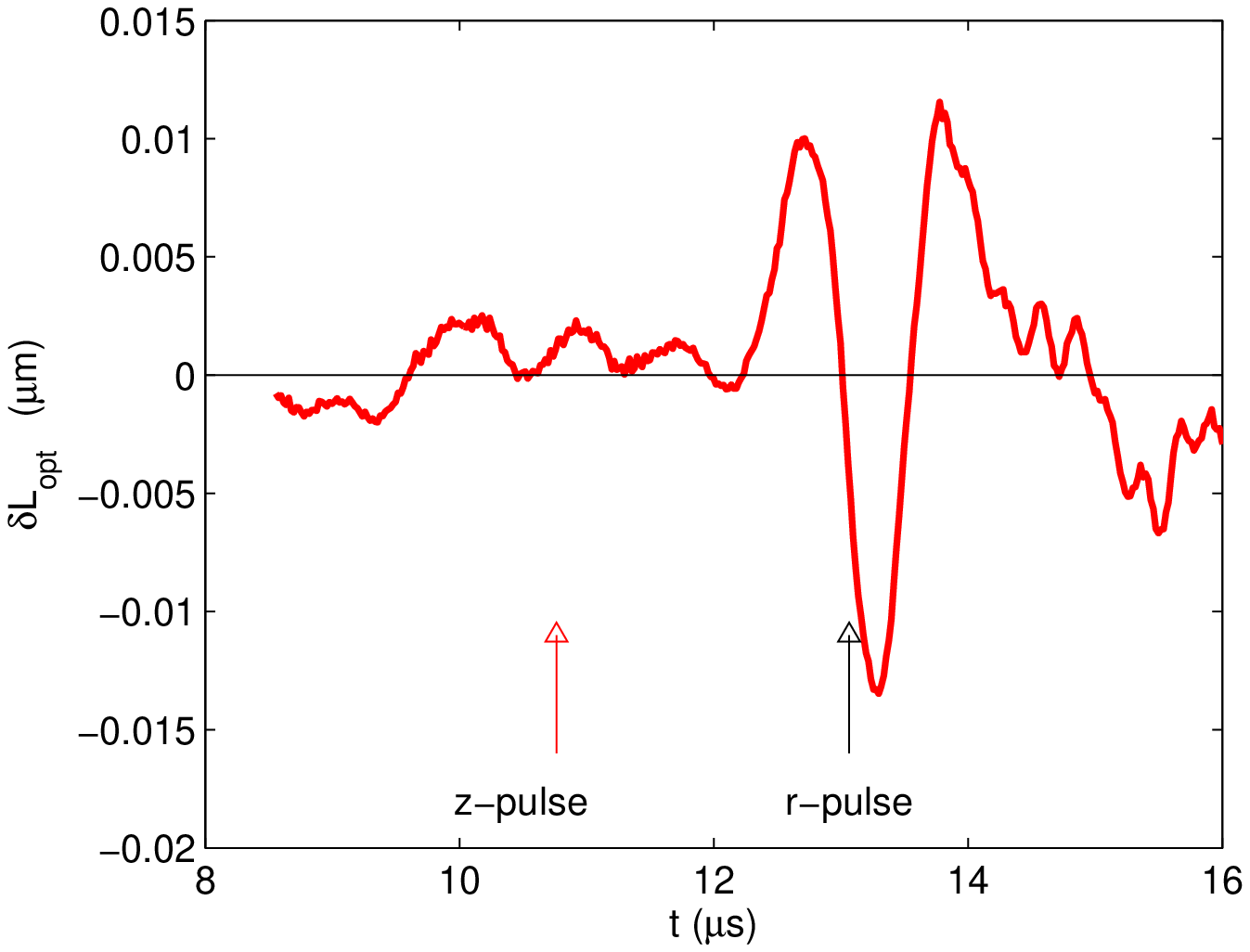} & \includegraphics[scale=0.38]{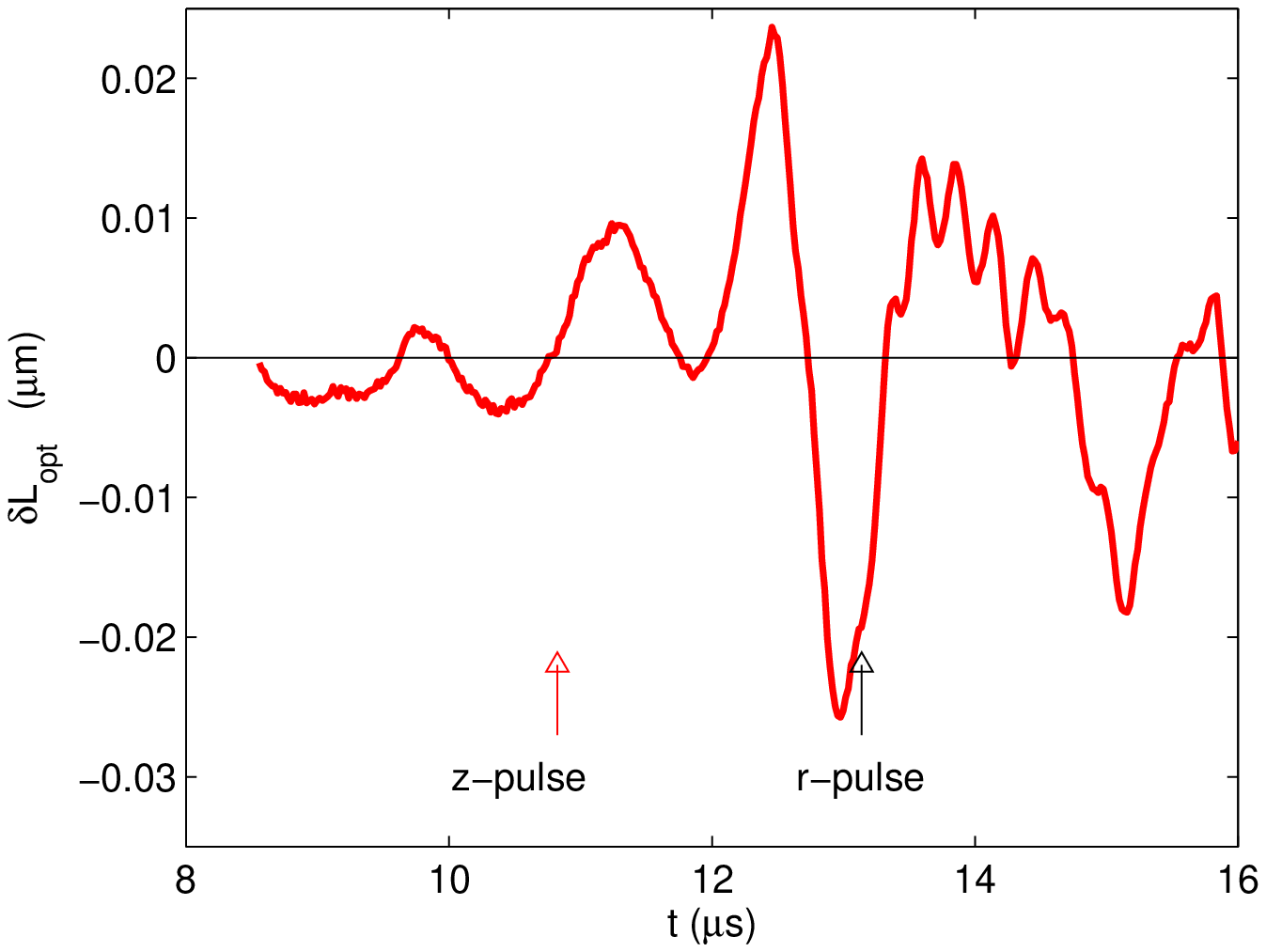}\\
(a)&(b)
\end{tabular}
}
\end{center}
\caption{Observed optical length modulation by  focused sound pulses. Time origin is the beginning of the transducer oscillation which lasts 1~$\mu$s.  Arrows indicate the middle of expected $z$- and $r$-pulses. (a) driving voltage 200~V, initial offset from melting pressure 0.6~bar, pressure swing at focus -0.35~bar. (b) Same signal with parameter values  600~V, 0.1~bar, -0.9~bar respectively.}\label{fig3}
\end{figure}

An example of observed optical signal $\delta L_{\rm opt}(t)$ is shown in Figure \ref{fig3}-a. It was taken with a driving voltage $V_d = 200$~V at  $T = 1.16$~K and a pressure 0.6~bar above $P_f(T)$. An order of magnitude of the sound amplitude at the transducer surface can evaluated from the formula $\delta\rho /\rho = 2\pi d_{33} V_d / \lambda_s f_r$, where $d_{33}=0.3$~nm/V  is the piezoelectric constant at room temperature of the transducer ceramic, and $f_r \simeq 4$ is an order of magnitude of  its reduction factor at low temperatures\cite{morganEC}. One finds $\delta\rho /\rho \simeq 2\times10^{-4}$, which can be converted to an acoustic pressure $\delta P=54$~mbar using solid helium bulk modulus  $B_s= 269$~bar.\cite{grilly1973}. Thus the minimum pressure at the transducer surface is well above the melting pressure. Note that for low sound amplitudes ($\delta\rho /\rho < 10^{-2}$), the computation of $\delta P$ using the bulk modulus instead of a non linear equation of state does not lead to a relative error larger than $10^{-2}$.

Comparison of the optical signal with Figure \ref{fig2}-b helps to identify  the radial wave. The expected time of flight is indicated by the arrow labeled \lq r-pulse\rq. Note that in Figures \ref{fig2} and \ref{fig3} the origin of time is the beginning of the transducer oscillation. To get the time of flight for the middle of the sound pulse, half its duration is to be subtracted, namely 0.5$\mu$s. The sound velocity at this particular pressure is determined using the scaling of $c_{11}$ and $c_{33}$ proposed by Maris\cite{maris2009}. For this signal, the laser beam  is thought to cross the focal region. In fact extensive scanning  of the focal region along the $x$ and $z$-axis with the laser beam was performed  to locate the sound focus. 

The shape of the $r$-pulse is qualitatively similar to the expected one. The width in time measured by the delay between the secondary maximums is  close to the 1.0~$\mu$s delay in Figure \ref{fig2}-b. There are several differences however. Some are parasitic effects. The high frequency modulation may be attributed to higher oscillation modes of the transducer which are found near 2.4~MHz and 3.6MHz. Oscillations before the expected $z$-pulse can be explained by the fact that the reference light beam is slightly perturbed by the wave emitted by the rim of the piezo-hemisphere.The distance between this rim and the reference beam is 4~mm and thus the sound wave reaches it in 7~$\mu$s. Concerning the $r$-pulse itself, the ratio of the main peak amplitude to the secondary maximums  is  much smaller than in the simulation. This could be due to an offset of the laser beam from the radial geometry or spatial averaging over the laser waist. Simulation shows that an offset of 0.1~mm would divide the peak by a factor 3. 

There is not a clear evidence for the $z$-pulse. It is either absent or hidden by parasitic oscillations. Its weakening may be correlated to that of the $r$-pulse peak. Another possible explanation is that the initially well oriented mono-crystal has been broken and turned to a more or less isotropic polycrystal by the sound wave. 
As a matter of fact, during long measurement series for scanning the laser beam position, we occasionally observed evolution of the optical signal, generally a weakening of the pulse amplitude.

Let us come back to the intensity of the negative swing at focus. Using the value of $a$ determined by the simulation ($a=0.28$~mm), one gets an order of magnitude of the negative pressure swing from the formula $\delta P =  \delta L_{\rm opt}  B_s / a (n-1) \simeq  -0.35$~bar.  The initial pressure was 0.6~bar from the melting pressure, so that the minimum pressure does not reach the melting line. 
Figure \ref{fig3}-b shows another signal taken with a larger driving voltage  (600 V) and starting closer to the melting line, $ P-P_f=0.11$~bar. The signal is clearly more distorted than in Figure  \ref{fig3}-a. If one converts nevertheless the negative swing into pressure using the same formula quoted above, one gets $\delta P =  -0.9$~bar. In that case, it appears that the pressure has crossed  the equilibrium melting line by at least 0.8~bar. 

Has helium remained in the solid phase? Had a liquid droplet nucleated at the focus, it would have expanded for a fraction of the sound half period at a velocity close to the sound velocity. Thus it would have reached a diameter on the order of 100~$\mu$m. Taking into account the refractive index  difference between liquid and solid phases (about 0.0035), the liquid bubble would have created a jump in $\delta L_{opt}$ as large as 0.35~$\mu$m. Also a large asymmetry would be found between positive and negative pressure swings. Neither of the two phenomena are observed and it can be concluded that transient melting did not occur. 
   
\section{Possible improvements and conclusions}
A direct interferometric imaging has  been developed \cite{souris2010}  to provide a map of  $\delta L_{\rm opt}$.  It uses the same interferometer, but the laser is a pulsed laser and the detector a CCD camera. Images are taken for various  $t$ and  reference phases. Then a map of $\delta\phi(t,x,z)$ can be computed for all the pixels in parallel. Beside providing a direct image of the sound  pulse propagation in the crystal, inverse Abel transform will give access to the map of    $\delta n(t,r,z)$, provided that it has a cylindrical symmetry. This should overcome the major shortcoming in the present experiment, namely the conversion from the measured $\delta L_{\rm opt}$ to the pressure swing $\delta P$ at the focus. Since the observed signal differs from the simulation, a more direct way to computed the latter will be valuable. It will also avoid scanning the laser beam to locate the focus, which is both time consuming and harmful for the crystal. In order to view the focus and an extended portion of the x-axis, the lower part of the hemispherical transducer will be shorten by 1~mm or so. This new method will also help assessing whether the crystal is broken by repetitive sound pulses. If this is the case, it would be of interest to try to work  with a polycrystal  with grains smaller than $\lambda_s$. The medium will then be quasi-isotropic\cite{maris2010}. Whether the sound attenuation will be acceptable remains an open question. In the case  monocrystals survive their repetitive stress, it would be more efficient to use a transducer with a shape conforming the wavesurface of the hcp crystal, or at least half of it. As depicted in reference \cite{crepeau1972}, it is an elongated bowl, nearly conical in the directions  between $z$ and $x$-axis. Finally using  longer multi-oscillation excitation pulses will allow to reach oscillation amplitudes $\delta P$ about 10 times larger. The more complex wave pattern will not be a problem if inverse Abel transform can be performed. 

To summarize, it has been shown that interesting motivations exist to investigate solid helium below the melting pressure. Numerical simulations confirmed that sound waves can be efficiently focused  in the anisotropic hcp crystal and provides a tool to investigate the metastable region below the melting curve.  A first example of incursion in this domain has been reported, and perspectives exist to enlarge the explored domain.

\begin{acknowledgements}
We acknowledge support from ANR, grant 05-BLAN-0084-01.
\end{acknowledgements}

\end{document}